\documentclass[prc,aps,floatfix,nofootinbib,twocolumn,letterpaper]{revtex4} 
\usepackage{graphicx} 
\usepackage{amssymb} 
\usepackage{amsmath} 
\usepackage{amsfonts} 

\usepackage{color}

\def\be{\begin{equation}} 
\def\ee{\end{equation}} 
\def\im{{\rm Im}}

\begin{document} 

\title{Transition-state dynamics in complex quantum systems}

\author{G.F. Bertsch}
\affiliation{ 
Department of Physics and Institute for Nuclear Theory, Box 351560, 
University of Washington, Seattle, Washington 98915, USA}

\author{K. Hagino}
\affiliation{ 
Department of Physics, Kyoto University, Kyoto 606-8502,  Japan} 

\begin{abstract}

A model is proposed for studying  the reaction dynamics in 
complex quantum systems in which the complete mixing of states is
hindered by an internal barrier.  Such systems are often treated
by the transition-state theory, also known in chemistry as RRKM theory,
but the validity of the theory is questionable when there 
is no identifiable coordinate associated with the barrier.
The model 
consists of two Gaussian Orthogonal Ensembles (GOE) of internal
levels coupled to each other and to the wave functions in the
entrance and decay channels. We find that
the transition-state formula can be derived from the model under
some easily justifiable approximations.
In particular, the assumption in transition-state
theory  that the reaction rates are 
insensitive to the decay widths of the internal states on the far
side of the barrier is fulfilled for broad range of Hamiltonian
parameters.  More doubtful is the common assumption that the transmission factor 
$T$ across the barrier is unity or can be modeled by a one-dimensional
Hamiltonian giving $T$ close to unity above the barrier.   This is not the
case in the model;  we find that the transmission factor only approaches one 
under special conditions that are not likely to be fulfilled without
a strong collective component in the Hamiltonian.  
\end{abstract}
 
\maketitle 
 
\section{Introduction}
The usual framework for describing quantum dynamics at a barrier crossing is
transition-state theory \cite{BW39,tgk96,thh83,MJ94,M74,LK83}, 
also known in chemistry as RRKM theory \cite{MR51,M52}.
The basic formula
of the theory gives the average decay rate   $\Gamma$ from one set of
states to another as
\be
\label{1}
\Gamma = \frac{1}{2\pi \rho} \sum_c T_c.
\ee
Here $\rho$ is the level density in the first set, $c$ labels a bridge channel
between the sets, and $T_c$ is the transmission factor through the channel.
As part of the theory, the transmission
factors are restricted to the range\footnote{In practice the
transmission factors are highly fluctuating as a function of energy;
the formula applies only to the average value.  These Porter-Thomas 
fluctuations \cite{th56} will be the subject of another article.} $ 0 < T_c < 1$.
This remarkable formula does not depend on the details of the Hamiltonian
outside the barrier region except for the level density in the first set of states. In 
particular, details of the Hamiltonian in the second set of states are irrelevant
except in determining the transmission factor;
we call this the {\it insensitivity} property. 
Finally, the formula is applied to reaction theory
without any need of information about the entrance channel coupling except
the total reaction cross section.

\begin{figure}[htb] 
\begin{center} 
\includegraphics[trim= 0   0  0  0, clip=true, 
width=\columnwidth]{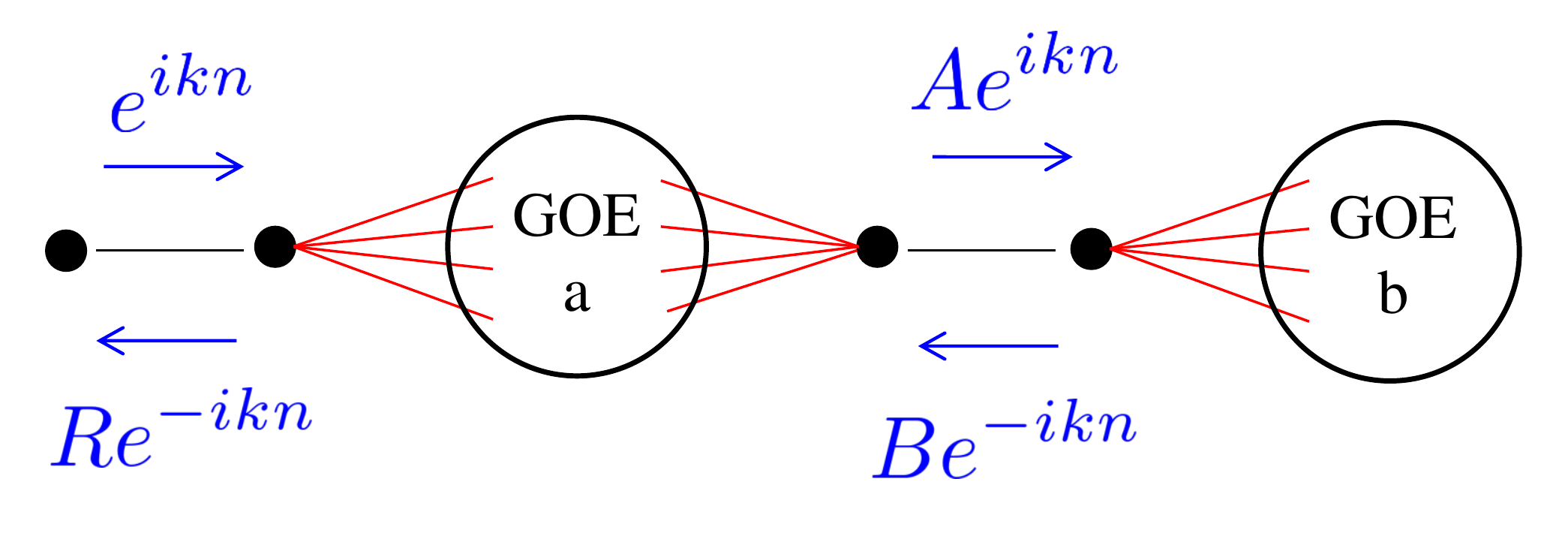} 
\caption{Connectivity of the Hamiltonian. The two dots on the left-hand
side represent states in discrete-basis representation of the entrance 
channel.  The dots between the GOE Hamiltonians define an internal
channel between them. The quantities in blue are the wave function
amplitudes separated into right-moving and left-moving parts, 
Eqs. (\ref{7})-(\ref{10}). }
\label{fig1} 
\end{center} 
\end{figure} 

In this article we construct a reaction model to test the theory by its
predictions for the branching ratio between decays from one set of states
or the other.  The Hamiltonian for each set will be taken from
the Gaussian Orthogonal Ensemble (GOE) \cite{W84,WM09,br81}.  
The wave function for the
channel spaces is usually treated by a formal separation of coordinates to
define a set of internal coordinates together with a continuous coordinate
governing the propagation in the channel.  Here we describe the channels in
a discrete-basis formalism \cite{al20}.  This is quite natural for the dynamics of
electrons in lattices, but it is also helpful to
construct channels involving complex fragments \cite{be19}.   The structure of the 
Hamiltonian is depicted in Fig. 1, with the 
entrance channel on the left-hand side and the bridge channel located between the
two GOE reservoirs.  We note that a two-reservoir model
with GOE ensembles has been used to study fluctuations of decay
widths in unimolecular reactions (see Ref. \cite{po90} and citations therein).

\section{Hamiltonian}

The model Hamiltonian is defined as 
\be
\label{H} 
{\mathbf H} = \left[\begin{matrix} 0 & t_1 & 0 & 0 & 0 & 0\cr
t_1 & 0 &  {\vec v}_2^T&0&0&0 \cr
0 &  {\vec v}_2 & H^{goe}_a - i\Gamma_a/2 &  {\vec v}_3^T & 0 & 0 \cr
0&0&\vec v_3 &0 &t_2 &0 \cr
0&0&0  & t_2 & 0 & {\vec v}_4^T\cr
0&0&0&0&\vec v_4 & H^{goe}_b-i\Gamma_b/2\cr
\end{matrix}\right].
\ee
The first 2-by-2 subblock applies to nearest-neigbor amplitudes in the discretized
entrance-channel wave function. Couplings to other states in the entrance
channel will be treated implicitly.  The next row and column applies to
the wave function in the 
first reservoir of dimension $N_a$. Next comes an internal channel
consisting of two coupled sites, followed
by the second reservoir of dimension $N_b$.  The total dimension of $\mathbf H$ is thus $2 + N_a + 2 + N_b$.
The parameters $t_1$ and $t_2$ are hopping matrix elements associated with
the channels.  The embedded GOE Hamiltonians have two parameters, the number of states
in the basis $N_g$ and the root-mean-square value $v_g$ of the 
matrix elements. They are computed as \cite{WM09,br81}
\be
\langle i|H^{goe}_g|j \rangle = r_{ij}  v_g (1 + \delta_{i,j} )^{1/2}
\ee
where $r_{ij}$ is a random number from a Gaussian distribution
of unit dispersion, $\langle r_{ij}^2 \rangle = 1$.  
The vectors $\vec v_k$ coupling the channels to the GOE states
are defined similarly, $\vec v_k(i) = r_i v_k$.  
Inelastic reactions proceed by decays through the states in the
two GOE sets.  The decays  from the two reservoirs to other
channels are
controlled by the widths $\Gamma_a$ and $\Gamma_b$. 

The main properties of the GOE spectrum needed in this work are
the near-Gaussian distribution of the eigenvector amplitudes
and the absence of correlations between them. 
As seen in Eq. (1), The derived properties depend crucially  
on the level density.  This is given by  Wigner's semicircular distribution, 
which we write as
\be
\rho(E) = \rho_0 \sqrt{ 1 - (E/E_m)^2}
\ee
where 
\be
\rho_0 = \frac{N_g^{1/2}}{\pi v_g}
\ee
is the level density at the center of the spectrum and
\be
 E_m= 2 N_g^{1/2}v_g
\ee
is the half-width of the spectrum.

\section{Reaction theory}

There are many formulations of reaction theory for complex
systems \cite{th09,ch13,ka15,be17,al20,da01}.  For the specific
form of our model Hamiltonian (2) it is straightforward to
deal directly with the wave function amplitudes and determine
the reaction rates by the associated fluxes.  To this end, it is convenient
to define an effective Hamiltonian that acts on the wave function
amplitudes $\phi_1,\phi_2$ of the entrance channel and the
amplitudes $\phi_3,\phi_4$ of the channel bridging the two GOE
sets of states.  The GOE states enter the effective Hamiltonian
via self-energy terms

\be
\label{wkk'-def}
w_{kk'} =  \vec v^T_k \cdot (E - H_g^{goe}+i \Gamma_g/2)^{-1} \cdot \vec
v_{k'}.
\ee
The effective Hamiltonian and the associated channel wave
function satisfy the equation
\be
\label{H2} 
\left[\begin{matrix} 
w_{22} - E &  w_{23} & 0 \cr
w_{23} &  w_{33}-E & t_2 \cr
0 & t_2 & w_{44}-E \cr
\end{matrix}\right]
\left(\begin{matrix} \phi_2 \cr \phi_3 \cr \phi_4 \cr
\end{matrix}\right) =
-\left(\begin{matrix} t_1\phi_1 \cr 0 \cr 0 \cr
\end{matrix}\right).
\ee
Here the GOE Hamiltonian  $H_a^{goe}$ goes into the calculation of $w_{22},w_{33}$, and $w_{23}$, 
while $w_{44}$ is calculated with $H_b^{goe}$.   
Eq. ({\ref{H2}}) is easily solved for the amplitudes on the left-hand side; details are
given in the Appendix A.
In the remainder of this article we will set $E=0$.  This places the
incoming energy at the centers of the GOE spectra. 

Following the procedure of Ref. \cite{al20}, the channel wave functions  are separated into right-going and left-going 
amplitudes as indicated in Fig. \ref{1}, 
\begin{align}
\label{7}
\phi_1 &= 1 - R\\
 \phi_2 &= e^{ik_1} - Re^{-ik_1}\\
\phi_3 &= A - B  \\
\phi_4 &= Ae^{ik_2} -B e^{-ik_2}.
\label{10}
\end{align}
The momentum index $k_i$ depends on $E$ and $t_i$ by the
relation $k_i = \cos^{-1}(E/2t_i)$; it is  $k_i=\pi/2$ at
$E=0$.

The inelastic cross sections  may be expressed by a kinematic factor times
the transmission coefficients $T$ from the entrance channel to a
set of decay channels.  The transmission 
coefficient into the system as a whole is 
\be
T = T_a + T_b = 1 -|R|^2
\ee
and the coefficient for the $b$ set alone is
\be
T_b = |A|^2 - |B|^2.
\ee 

The detailed formulas for the reflection coefficient $R$ and the
bridge amplitudes  $A,B$ are given in the Appendix A.

\section{Branching ratios}

The physical observable we examine in this article is
the branching ratio $B_r$ between the two sets of exit channels.  This may be
computed from the reaction amplitudes as
\be
B_r = \frac{T_b}{T_a} = \frac{|A|^2 - |B|^2}{1-|R|^2 -|A|^2 + |B|^2}
\ee 
or by calculating the fluxes directly.
The flux  into the
system from the entrance channel 
is given by 
\be
\Phi_{12} = 2 t_1 {\rm Im} ( \phi_1 \phi^*_2)
\ee 
and the flux into
the second reservoir is given by the corresponding expression
with $\phi_3,\phi_4$ and $t_2$.  
The branching ratio is thus
\be
B_r = \frac{\Phi_{34}}{\Phi_{12}- \Phi_{34}}.
\ee
In terms of the parameters in the effective Hamiltonian the
branching ratio is
\be
\label{Br_full}
B_r = 
  \frac{t_2^2 |w_{23}|^2 {\rm Im}(w_{44})}
{{\rm Im}(w_{22}) |s|^2  -
\im(w_{23}^2 w_{44} s^*) - t_2^2 |w_{23}|^2  \im(w_{44})}, 
\ee
where
\be
s = w_{33} w_{44} - t_2^2.
\ee

Note that Eq. (\ref{Br_full}) for the branching ratio is independent of the parameters that control the
coupling to the entrance channel.  The parameter $t_1$ does not
appear in the formula at all.  The coupling parameter $v_2$ is
implicit in the definitions of $w_{22}$ and $w_{23}$, but
the numerator and denominator are both proportional to $v_2^2$
and cancel out.  The independence of entrance channel
details is not surprising;  the entrance channel
populates the states in reservoir $a$, but after that only
the internal Hamiltonian affects the decay.  

\section{Mean values of the self-energies}
\label{V}
The physics of the Hamiltonian is buried in 
the four complex numbers  $w_{ij}$ in Eq. (\ref{H2}).
In this section we derive their average values.  We
first consider the diagonal ones $w_{kk}$ and then the off-diagonal
one $w_{kk'} = w_{23}$.

\subsection{Statistics of the diagonal $w_{kk}$}
In this section we will estimate the self-energies in the effective
Hamiltonian by taking a continuum limit over the sum over states in
the Green's function $G(E)= (E - H_g^{goe}+i \Gamma_g/2)^{-1}$.  This Green's
function has been analyzed in much more detail in Ref. \cite{de07} for
different purposes.

The elements of the Green's function can be written in the eigenstate 
representation as
\be
\label{Geig}
G_{ij} = \sum_\lambda^{N_g} \frac{\phi_\lambda(i)\phi_\lambda(j)}{E -
E_\lambda}.  
\ee
Here $E_\lambda$ are the eigenvalues of the GOE Hamiltonian including the
imaginary offset $-i\Gamma_g/2$,  and the $\phi_\lambda$ are the corresponding
eigenvectors. In general the eigenvectors of a complex matrix are
necessarily complex as well, but since the imaginary part of the Hamiltonian
is a constant offset we can choose them to be real.   

We next replace the sum in Eq. (\ref{Geig}) by an integral over the level
density. See Appendix B for a list of integrals needed in this section. 
As a simple example, the Stieltjes transform of Tr$(G)$ in 
dimensionless form \cite{pa72} is given by the integral $I_1$ in the
appendix,  
\be
\label{stieltjes}
f(z) = I_1(z) \equiv \int^{+1}_{-1} dx \frac{\sqrt{1 - x^2}}{z-x} = 
-\pi\left(\sqrt{z^2 - 1}-z\right). 
\ee
The replacement of the sum by an integral is most easily justified by
demanding that the width satisfies $\Gamma_g \gg \rho_0^{-1}$ but we
shall see later that this condition can be relaxed for ensemble
averages.  

\begin{figure}[tb] 
\includegraphics[width=\columnwidth]{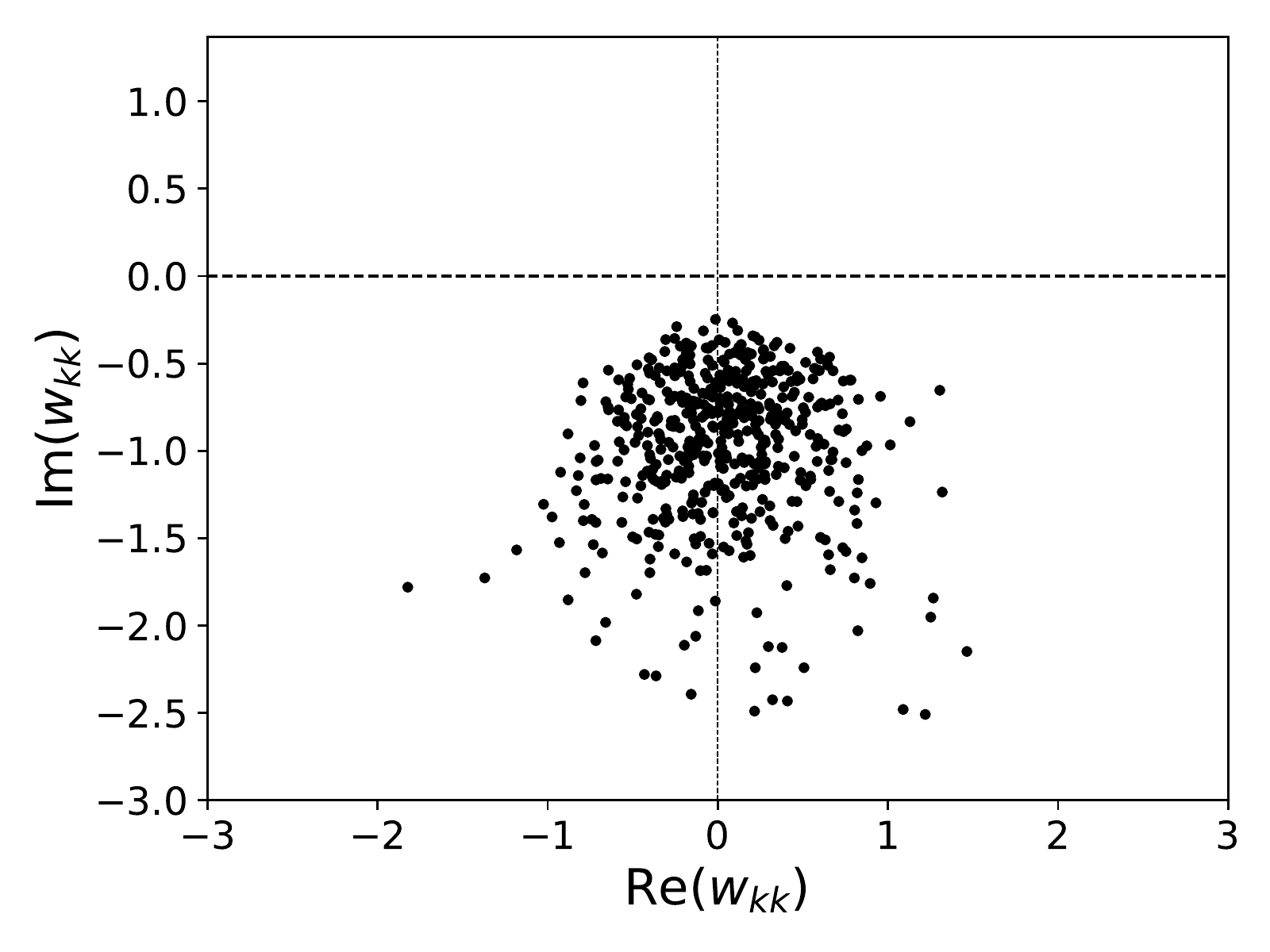} 
\caption{
Scatter plot of 500 samples of the $w_{kk}$ ensemble.
The Hamiltonian parameters are $N_g=100$ and $v_g,v_k,\Gamma_g =
 0.1$.
\label{wkk-scat} 
}
\end{figure}

\begin{figure}[tb] 
\includegraphics[width=\columnwidth]{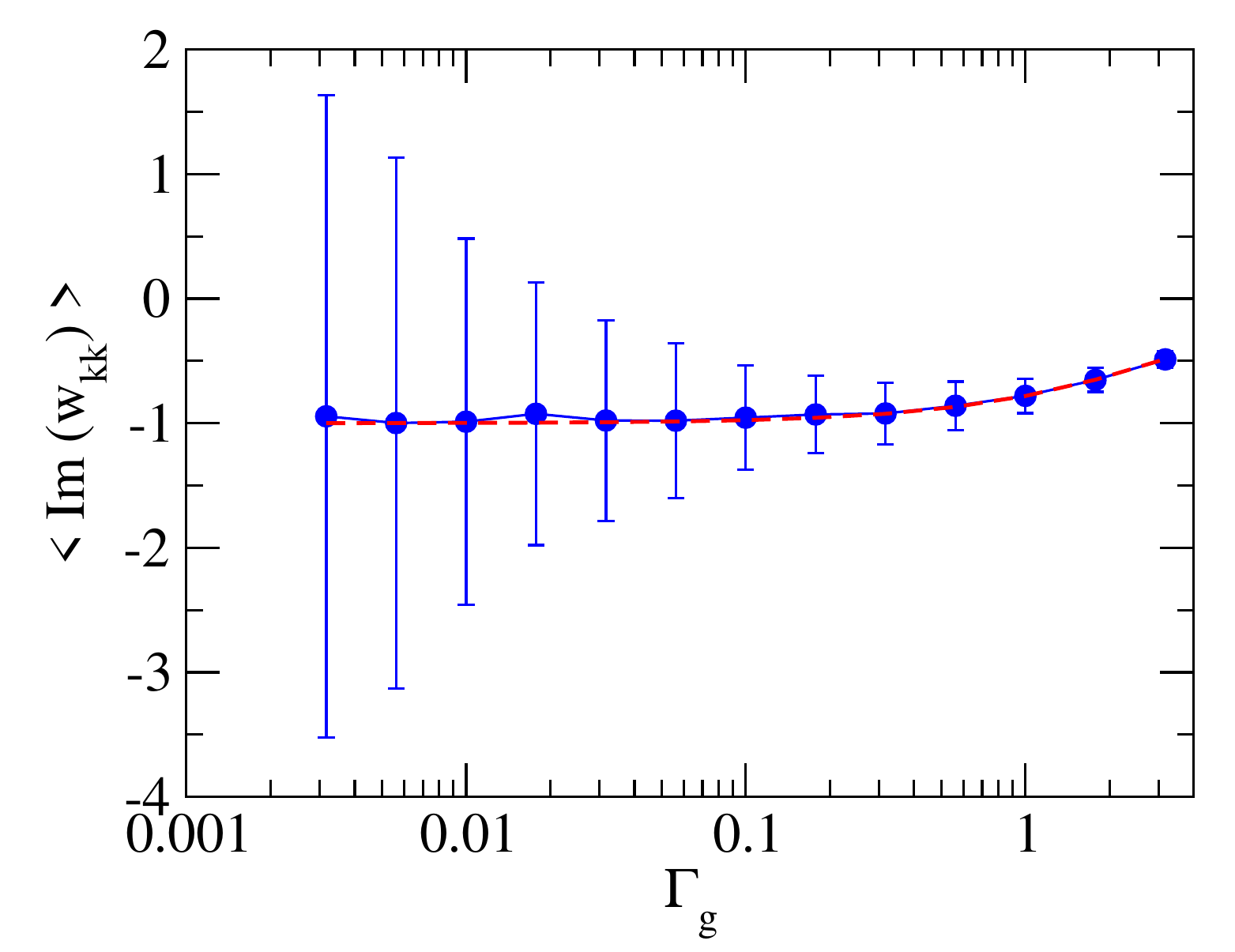} 
\caption{
Comparison of Eq.  (\ref{wkk-ans1}) with Eq.  (\ref{wkk-def}). 
The parameters are $N_g=100$, $v_g= 0.1$, and $v_k = 0.1$.  Eq. 
(\ref{wkk-def}) was evaluated statistically with 500 samples. 
The means are shown as filled circles, and the error bars
denote standard deviation of the samples about the means. 
The dashed line shows the corresponding value from Eq. 
(\ref{wkk-ans1}).
\label{wkk-fig} 
}
\end{figure}

The diagonal self-energy $w_{kk}$ can be expressed 
\be
\label{wkk-def}
w_{kk}(E) = \sum_\lambda^{N_g} \frac{(\vec v_k\cdot \vec \phi_\lambda)^2}
{E - E_\lambda}.
\ee
Fig. \ref{wkk-scat} shows a sampling of the ensemble for a set of Hamiltonian parameters
in a GOE space of dimension $N_g = 100$ at energy $E=0$.  There is
a clustering around a point on the negative imaginary axis, and the spread
of values is about the same in the real and imaginary directions.  To
find the clustering center we examine the ensemble average of
Eq. (\ref{wkk-def}),
\begin{align}
\label{wkk_ave}
w^e_{kk} &\equiv \langle w_{kk}(0) \rangle =  
\left\langle\sum_\lambda^{N_g} \frac{(\vec v_k\cdot \vec \phi_\lambda)^2}
{-E_\lambda}\right\rangle\\
  &= \langle(\vec v_k\cdot \vec \phi_\lambda)^2\rangle
\left\langle\sum_{\lambda'}^{Ng} \frac{1}{-E_\lambda}\right\rangle.
\end{align}
The second line follows from the fact that the eigenvector components are
uncorrelated with each other or the eigenenergies.
The sum is the same as in the formula for Tr$G$; its ensemble average
is 
\be
\left\langle\sum_{\lambda'}^{Ng} \frac{1}{-E_\lambda}\right\rangle =
-i \rho_0 I_1(z_g)  
\ee 
with $z_g = i\Gamma_g/2 E_m$.  
The dot products $(\vec v_k\cdot \vec \phi_\lambda)$ are Gaussian
distributed and the variance is easily seen to be 
\be
\langle (\vec v_k\cdot \vec \phi_j)^2 \rangle =  v_k^2.
\ee
Bringing these ingredients together, $w_{kk}^e$  is given by
\be
\label{wkk-ans1}
w^e_{kk} =  -i v_k^2 \rho_0 I_1(z_0).
\ee
In physical applications, the widths of the states in the GOE will
be much smaller than the overall spread in the GOE, i.e. $|z_0|
\ll 1$.  The corresponding
limit in Appendix B yields
\be 
\label{wkk-ans2}
w^e_{kk} \approx -i \pi v_k^2 \rho_0\,\,\,{\rm for}\,\,\, \Gamma_g \ll E_m. 
\ee
Fig. \ref{wkk-fig} shows a graph of ${\rm Im}(w_{kk}^e)$
as function of $\Gamma_g$ 
obtained by Eq. (\ref{wkk-ans1}) and compared with a value 
direct sampling of Eq. (\ref{wkk-def}).  One sees
that the formula is in satisfactory agreement over the entire range
examined.  This is despite the fact that the fluctuations of the individual
samples are small only when $\Gamma_g \gg \rho_0^{-1}$.
An important consequence of Eq. (\ref{wkk-ans2}) is that $w_{kk}$ is 
quite insensitive to the width parameter $\Gamma_g$. 
Since the average reaction observables only depend on the decay width of
the second GOE set through $w_{44}$, this proves the insensitivity 
property.  Namely, when the system crosses the transition region
it remains in the farther set of states until the final decay into other
channels.  
Eq. (\ref{wkk-ans2})
can also be easily derived  from Fermi's Golden Rule, as noted in Ref.
\cite{de07}.

\begin{table}[htb] 
\begin{center} 
\begin{tabular}{|c|c|c|cc|} 
\hline 
\hline 
$N_g$ & Type &  sampled   &  formula & Eq. \\
\hline
100& $w_{kk}$ &   $-(0.02\pm 0.48) -(1.02\pm0.47)i$ & $-i$  &
(\ref{wkk-ans2})  \\
&$w_{kk'}$ &  $(0.03\pm0.32)+(0.06\pm0.35)i$ &   0  & \\
&$|w_{kk'}|^2$ & $0.23 \pm 0.27$  &  $0.2$ & (\ref{wkkp-phys}) \\
&$w_{kk'}^2$ & $(-0.03\pm0.21)+(0.02\pm0.27)i$ & $-0.005 $ &
(\ref{wkkp2-phys}) \\
\hline 
400 &$w_{kk}$ &   $-(0.00025\pm 0.61) -(1.95\pm0.55)i$ & $-2i$  &
(\ref{wkk-ans2})  \\
&$w_{kk'}$ &  $(-0.014\pm0.44)+(0.0063\pm0.47)i$ &   0  & \\
&$|w_{kk'}|^2$ & $0.41 \pm 0.41$  &  $0.4$ & (\ref{wkkp-phys}) \\
&$w_{kk'}^2$ & $(-0.031\pm0.44)+(0.045\pm0.38)i$ & $-0.005 $ &
(\ref{wkkp2-phys}) \\
\hline 
\hline 
\end{tabular} 
\caption{
Expectation values and statistical fluctuations of self-energies associated
with the coupling of channels to GOE ensembles with the dimension $N_g$. 
The energy parameters
$v_g,v_k,v_{k'},\Gamma_g$ are set to 0.1.  
The sampled results in the third 
column are the mean values and rms fluctuations obtained from 100 independent
samples.
\label{wkk-table}
} 
\end{center} 
\end{table} 

\subsection{Statistics of the off-diagonal $w_{kk'}$}
\begin{figure}[tb] 
\includegraphics[width=\columnwidth]{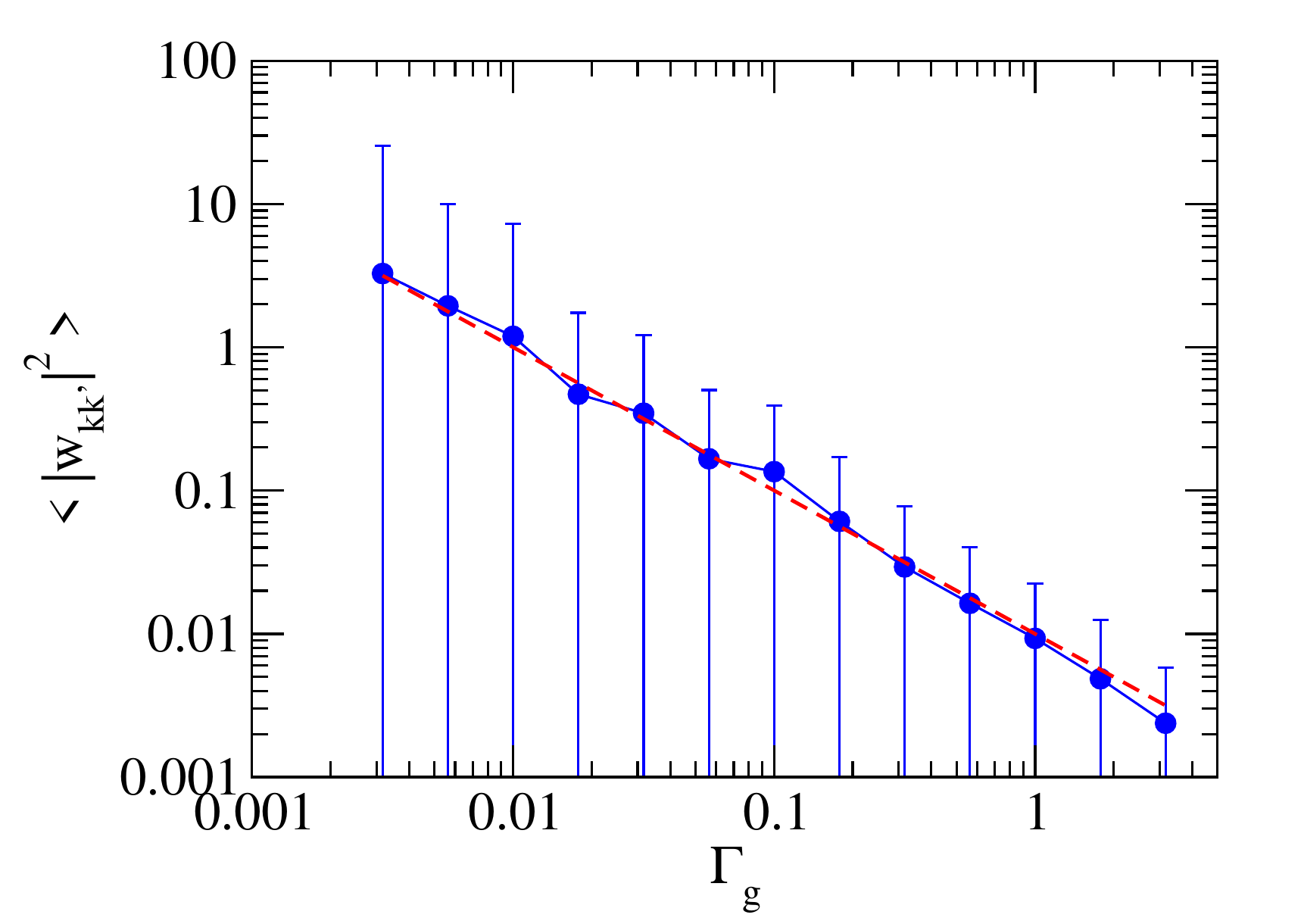} 
\caption{The self-energy term $\langle |w_{kk'}|^2\rangle$ comparing the formula Eq.
(\ref{wkkp-phys}) with the  mean value evaluated with 500 samples from the
Hamiltonian ensemble. The mean is shown with filled circles.  
The errors denote the r.m.s. deviation of the samples about the
mean.  The dashed line shows the corresponding value from Eq. (\ref{wkkp-phys}).
\label{wkk'-fig} 
}
\end{figure}

The mean value of the off-diagonal $w_{kk'}$ is zero since
the  external coupling vectors $\vec v_k$ and
$\vec v_{k'}$ are independent.  However, the branching ratio in Eq. (\ref{Br_full}) 
requires only the squared quantities $w_{kk'}^2$  and $|w_{kk'}|^2$.
To determine the expectation values of these quantities, we first express them
in term of the eigenvalues and eigenfunctions of $H_g^{goe}
-i \Gamma_g/2$.  The expression for $|w_{kk'}|^2$
is
\be
\label{wkkpasq}
|w_{kk'}(E)|^2  = \sum_{\lambda,\lambda'}^{N_g} \frac{u_{kk'}(\lambda)
u_{kk'}(\lambda')}
{(E - E_\lambda)(E - E_{\lambda'}^*)}  
\ee 
where $ u_{kk'}(\lambda) = (\vec v_k\cdot \vec \phi_\lambda)
(\vec v_{k'}\cdot \vec \phi_{\lambda})$. As before, we expose the
statistical properties of the dot products by writing them as
\be
\vec v_k\cdot \vec \phi_\lambda = v_k\, r,
\ee
where $r$ again is a Gaussian variable with variance $ \langle r^2 \rangle = 1$.
Thus we can write $u_{kk'}(\lambda) = v_k v_{k'} r r'$  where $r$ and $r'$
are independent.  The double sum 
in Eq. (\ref{wkkpasq}) reduces to a single sum because the $u$'s for 
different $\lambda$'s are uncorrelated.  The numerator in the reduced 
sum is given by 
\be
u_{kk'}(\lambda)^2 = v_k^2 v_{k'}^2 r^2 r'^2,
\ee    
which has an expectation value
\be
\langle u_{kk'}(\lambda)^2 \rangle  = v_k^2 v_{k'}^2.
\ee
The remaining task is to do the sum over $\lambda$, which we again approximate
as an integral.  Setting $E=0$, the expectation value of $|w_{kk'}|^2$ becomes
\begin{align}
\langle|w_{kk'}(0)|^2\rangle  &=  v_k^2 v_{k'}^2 \frac{\rho_0}{E_m} I_3(c) \label{wkkpasq-phys}
\\
   & \approx  2\pi v_k^2v_{k'}^2 \frac{\rho_0}{\Gamma_g} \label{wkkp-phys}
\end{align}
where $I_3$ is the integral in Eq.(B3) and its argument is $c = \Gamma_g/2
E_m$.  In the last line the integral has been evaluated approximately for
$c \ll 1$.
Fig. \ref{wkk'-fig} shows 
$\langle |w_{kk'}(0)|^2\rangle$ 
from direct sampling of Eq. (\ref{wkk-def}). The agreement with 
Eq. (\ref{wkkp-phys}) is remarkable.

The statistical properties of $|w_{kk'}|^2$ for some  numerical 
examples are shown in Table I (third row in each subtable).  One can see that the mean
values satisfy   Eq. (\ref{wkkp-phys}) fairly well, even though there is a large 
fluctuation.  

The expectation value of $w_{kk'}^2$  can be evaluated in a similar way
with the integral $I_2$ in Appendix B.
In the physically interesting region  ($ c \ll 1$) it becomes
\be
\label{wkkp2-phys}
\langle w_{kk'}^2\rangle \approx   -\pi v_2^2
v_3^2\frac{\rho_0}{E_m}.
\ee
This is smaller than $|w_{23}|^2$ by a factor of $\Gamma_g /2 E_m$ and
can be ignored in deriving the transition-state limit below.

\section{Transition-state limit} 
The transition-state formula Eq. (1) as applied to a single channel
can now be derived from Eq. (\ref{Br_full}) under certain conditions.  
We neglect the
second and third terms in the denominator in Eq. (\ref{Br_full}) and write
\be
\label{br-approx}
B_r \approx
  \frac{t_2^2 |w_{23}|^2 {\rm Im}(w_{44})}
{{\rm Im}(w_{22}) |w_{33} w_{44} - t_2^2|^2}.
\ee
Next, we define an average value $\overline B_r$ by replacing the fluctuating quantities in the
above equation by their expectation values.  This is an approximation
because it neglects correlations between them.  The result is
\begin{eqnarray}
B_r &\approx& \frac{1}{2\pi \rho_0 \Gamma_a}
\frac{4 N_a t_2^2 v_a^2 v_3^2 v_4^2}{|v_3^3 v_4^2 N_a + t_2^2 v_a^2|^2} \\
&=&\left(\frac{1}{2\pi \rho_0 \Gamma_a}\right)
\left(\,\frac{-4t_2^2w_{33}w_{44}}{|-w_{33}w_{44}+t_2^2|^2}\right).
\label{br_approx2}
\end{eqnarray}
The first term in parentheses is exactly the transition-state branching
ratio for $T=1$.  The second term in parentheses can be interpreted as
the transmission factor $T_{ab}$ between the two reservoirs.  The full 
derivation is given in Appendix C.  It is easy to show that it is
equal to one if the transmission factors between the bridge
channel and the two reservoirs are both unity, as may be seen by
the following arguments.  
The transmission factor
between the bridge channel and reservoir $b$  is 
\be
T_{cb} =  (|A|^2-|B|^2)/|A|^2
\ee
which reaches one when $|B|^2 = 0$. According to Eq. (\ref{B}), this
is achieved if $ w_{44} = -it_2$.    
A similar condition applies to the
transmission coefficient to the first reservoir,  
$w_{33} = -it_2$.  Imposing both conditions, the second factor in
Eq.(\ref{br_approx2}) reduces to unity.  

We have also evaluated the transition state formula numerically, sampling the
ensemble with Eq. (\ref{Br_full}).  Taking the example  
with all $v$'s and $\Gamma$'s equal to 0.1 and $t_1 = t_2 = -1$, the
self-energies  satisfy
the condition that $T_{ca}=1$ and $T_{cb}=1$. We obtain
\be
\label{br-num}
B_r = 0.048\pm0.048 \,\,\,{\rm (numerical)}
\ee  
by sampling the Hamiltonian
ensemble.  

\begin{figure}[tb] 
\includegraphics[clip=true,width=\columnwidth]{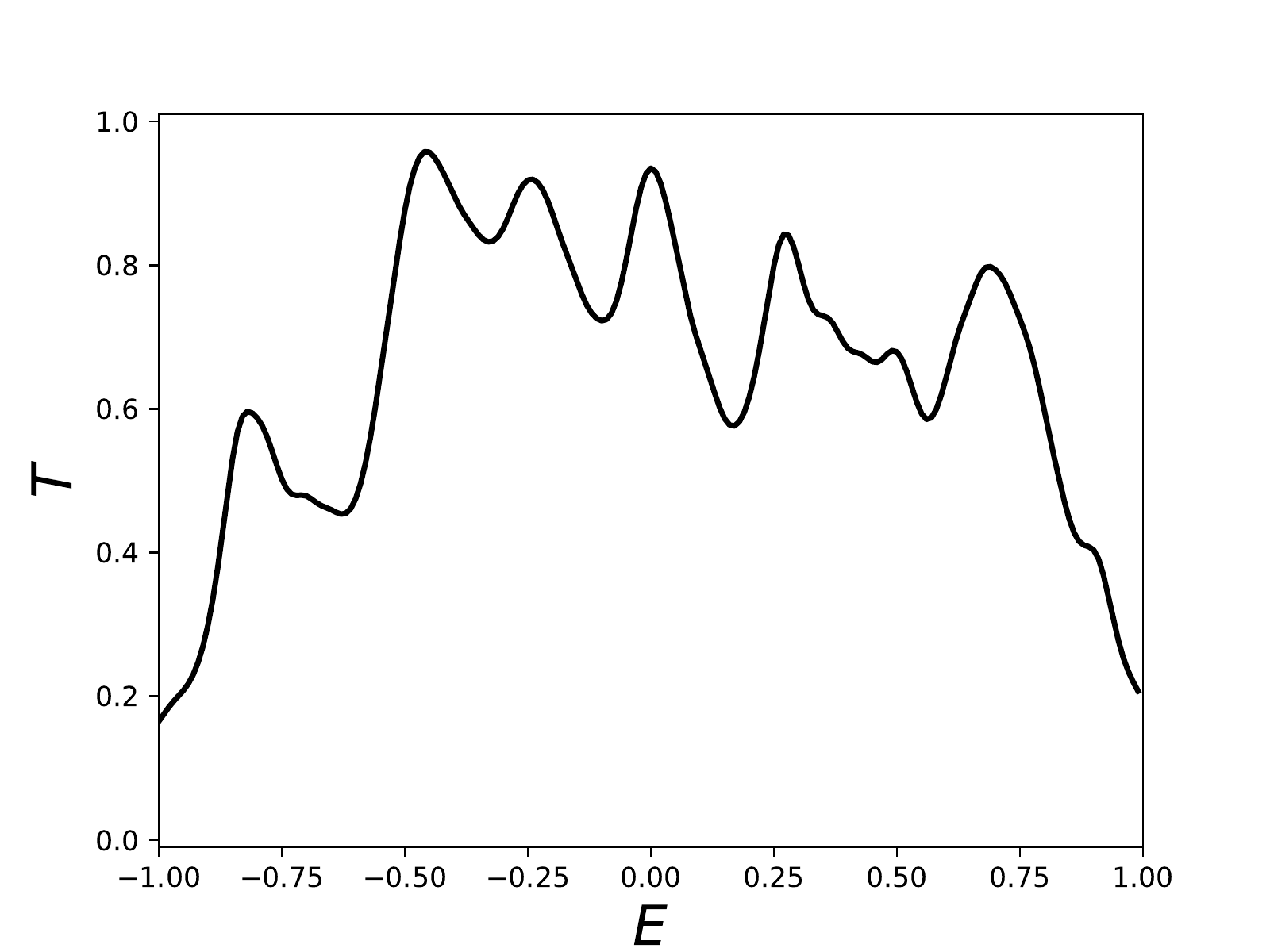} 
\caption{Transmission coefficient from a channel into a representative GOE
sample.  The GOE Hamiltonian parameters are the same as in Fig.
\ref{wkk-scat} and $t =-1$ in the channel Hamiltonian.
\label{T} 
}
\end{figure}

This is to be compared with the analytic formula which
gives 
\be
\overline B_r = 0.048\,\,\,{\rm (analytic)}.
\ee
The agreement here is somewhat unexpected, 
considering the fact that 
the averaged internal transmission factor can only be smaller
than one.  This may be seen in Fig. \ref{T} showing $T_{cb}$
as a function of $E$ for the same Hamiltonian as before. 

\section{Summary and Discussion}

The model described in Section II treats reactions involving a 
quantum system that can decay in two distinct ways, one through
states directly coupled to the entrance channel and the other through
a second set of states coupled to the first via an
internal channel (often called a ``transition state").  The states
in both sets are chaotically mixed as given by the well-known
Gaussian orthogonal ensemble.  The statistical properties of the
model can be analyzed to a large extent analytically, permitting a
qualified derivation of the  transition-state theory encapsulated
in  Eq. (1).
Two of our findings appear to be quite robust:

1) The branching ratio between the decay modes is independent of the
details of the coupling to the entrance channel.  This is important
if the reaction proceeds through isolated resonances visible from
the entrance channel---one does not have to deal with specific energies
of physical resonances to calculate average properties.

2) The transition-state formula does not require knowledge of the
decay widths of states in the second set, and this insensitivity
is verified over a large range of Hamiltonian parameters.
In other words, once the
flux has passed over the barrier, it does not matter how long 
it takes to decay from the states on the other side. The only
caveat is that the width should not be larger than the full
spread of the GOE spectrum.  
 
Our  derivation of the transition-state theory provides a formula
for the transmission coefficient in terms of the parameters
in the Hamiltonian ensemble.  A related finding
is:

3) Systems having a transmission factor  $T=1$ should be a rare 
occurrence, since it demands a specific coupling strength  
of the bridge channel to both sets of GOE states.
Furthermore, the coupling $t_2$ within the channel has to be much larger than the other interactions
in the Hamiltonian.  This requires some additional physics: in chemistry
this is provided by the kinetics of nuclear motion.  In Fermi systems it
might be provided by a superfluidity from a pairing interaction.

For a possible future application of the methods discussed here, one 
could consider two chaotic sets of states coupled through a single state instead
of a channel.  The expected transmission coefficient is given 
by the quantum-dot formula \cite[Eq. (21)]{al00},  
\be
T_{ab} = \frac{ \Gamma_a \Gamma_b}{(E - E_r)^2 + (\Gamma_a + \Gamma_b)^2/4}.
\ee
Here $E_r$ is the diagonal energy of the bridge state;  $\Gamma_a,\Gamma_b$
correspond to $\im(w_{33}),\im(w_{44})$ in our notation.

Some caveats in our analysis should be repeated.
First, our derivations of analytic formulas are valid only for
$\Gamma \gg \rho_0$, corresponding to the overlapping
resonance limit in the two sets of states. However, the statistical
properties of the self-energies and other derived quantities
such as ratio formula Eq. (\ref{Br_full}) appears to be valid also when the
resonances are isolated.  It would be
interesting to see if the derivations could be extended
to cover such cases as well. Second, we have ignored fluctuations
in the spectral function in calculating the statistical properties
of the GOE Green's function.  This may well be justified by 
Dyson's spectral rigidity, but our methodology is not adequate
to address that approximation.

In this paper we have concentrated on the mean values of the 
self-energies and the branching ratios. We will discuss their fluctuations, 
in particular, the validity of the Porter-Thomas distribution, 
in a separate publication \cite{bh2021-2}. 

\section*{Acknowledgments}
This work was supported in part by
JSPS KAKENHI Grant Number JP19K03861.

\appendix

\section{Reflection coefficient and the bridge amplitudes}
The wave function amplitudes $\phi_2,~\phi_3$, and $\phi_4$ obtained by inverting
the matrix in Eq. (\ref{H2}) are
\begin{align}
\phi_2 &= -(w_{33} w_{44} -t_2^2) t_1 \phi_1 /D \\
\phi_3 &= w_{23} w_{44} t_1 \phi_1 / D\\
\phi_4 &= -w_{23} t_2 t_1 \phi_1 / D
\end{align}
where
\be
D = w_{22} w_{33}w_{44} - w_{22} t_2^2 - w_{23}^2 w_{44}.
\ee

These expressions are inserted into Eq. (\ref{7}-\ref{10}) and solved
for the transport amplitudes $R,A,$ and $B$. The result at $E=0$ is
\begin{align}
R &= \frac{(w_{33} w_{44} - t_2^2)t_1 + iD}{(w_{33} w_{44} - t_2^2)t_1 
-iD}\\
A &= -\frac{t_1 w_{23}}{2 i D}
(1-R)(t_2 -i w_{44})\\
B &= -\frac{t_1 w_{23}}{2iD}
(1-R)(t_2 +i w_{44}).\label{B}
\end{align}

\begin{figure}[tb]
\includegraphics[width=\columnwidth]{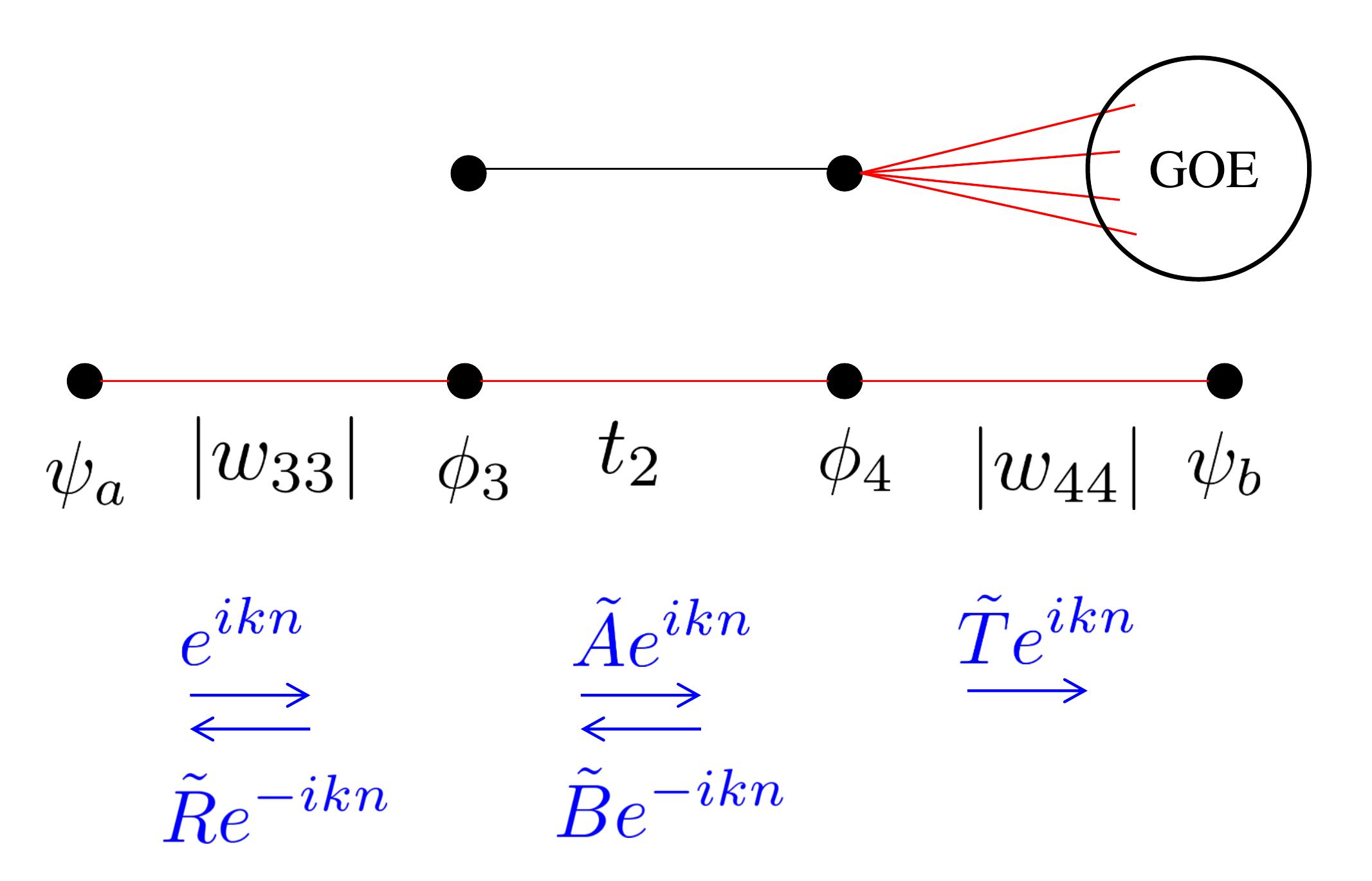}
\caption{
The chain of four states shows channels and the wave function amplitudes 
in the discrete-basis representation needed to evaluate 
the transmission factor $T_{ab}$.  The partial
transmission factors $T_{ca}$ and $T_{cb}$ are
set by the properties of the Hamiltonian consisting
of a channel with a single GOE reservoir, shown
at the top.     
\label{fig:appendix}
}
\end{figure}

\section{Self-energy integrals}

The statistical properties of the self-energy are calculated making
use of the following integrals:
\be
\label{stieltjes2}
 I_1(z) =  \int^{+1}_{-1} dx \frac{\sqrt{1-x^2}}{z-x} = 
\pi\left(z- \sqrt{z^2 - 1}\right) 
\ee
\begin{align}
I_2(z) =\int_{-1}^{+1} dx \frac{\sqrt{(1-x^2)}}{( z - x)^2} 
   =\pi \left(\frac{z}{(z^2-1)^{1/2}}-1\right) 
\end{align}
\begin{align}
I_3(y) = \int_{-1}^{+1} dx \frac{(1-x^2)^{1/2}}{(x^2 + y^2)}
   =\frac{\pi}{y}\left(\sqrt{1+y^2}-y\right)
\end{align}
\begin{align}
I_4(y) = \int_{-1}^{+1} dx \frac{(1-x^2)^{1/2}}{(x^2 + y^2)^2}
   =\frac{\pi }{2 y^3 \sqrt{1+y^2}}.
\end{align}

The first integral is the Stieltjes transform of the semicircular
GOE level density in dimensionless form as discussed by Pastur \cite{pa72}.  
The integral $I_2$ can be derived from it by differentiation. The integrals
$I_3$ and $I_3$ have been checked by numerical integration.

We especially require the leading behavior of the integrals for small
positive imaginary $z$ and small real $y$.  In those limits the integrals become
\be
I_1(z) \approx -i\pi
\ee   
\be
I_2(z) \approx -\pi   
\ee   
\be
I_3(y) \approx \frac{\pi}{y}   
\ee   
\be
I_4(y) \approx \frac{\pi}{2 y^3  }. 
\ee   

\section{Transmission factor}

Under certain conditions, the second  factor in Eq. (\ref{br_approx2}) 
reduces to the transmission factor $T_{ab}$ between the two reservoirs. To
derive $T_{ab}$ from the Hamiltonian parameters we need to know the
transmission and reflection amplitudes at the interface between the
bridge channel and both reservoirs.
The reflection probability at the interface between the bridge channel and the 
second reservoir is readily available as the ratio
\be
R = B/A = \frac{t_2+iw_{44}}{t_2-iw_{44}}.  
\ee  
Taking $w_{44}$ to be negative imaginary as in Eq. (\ref{wkk-ans2}), 
one can show that
the reflection amplitude would be the same for an interface to another
channel consisting of a chain of states connected by  nearest-neighbor 
interactions of magnitude $t_b = \im(w^e_{44})$  The same argument applies 
to the interface with the first reservoir, where the magnitude of the
coupling in the fictitious channel is  $t_a = \im(w^e_{33})$.

The surrogate Hamiltonian is depicted as the middle chain of four
states in Fig. \ref{fig:appendix}.  The two bridge states in the middle
are connected to the closest states in each channel.  The 
Hamiltonian equation for amplitudes $\phi_3$ and $\phi_4$ are given by
\begin{eqnarray}
&&|w_{33}|\psi_a+t_2\phi_4=E\phi_3=0, 
\label{eq:appendix1}
\\
&&t_2\phi_3+|w_{44}|\psi_b=E\phi_4=0, 
\label{eq:appendix2}
\end{eqnarray}
for scattering at $E=0$. 
The amplitudes on the
four sites are expressed in terms of the traveling wave amplitudes
$\tilde{R},\tilde{A},\tilde{B}$ and $\tilde{T}$ as shown in the figure.  The phase of these amplitudes
change by $e^{\pm i k}$ from one site to the next in a channel. These lead 
to the equations for the amplitudes
\begin{eqnarray}
\psi_a&=&e^{-3ik/2}-\tilde{R}e^{3ik/2}, \\
\phi_3&=&e^{-ik/2}-\tilde{R}e^{ik/2}
=\tilde{A}e^{-ik/2}+\tilde{B}e^{ik/2}, \\
\phi_4&=&\tilde{A}e^{ik/2}+\tilde{B}e^{-ik/2}=\tilde{T}e^{ik/2}, \\
\psi_b&=&\tilde{T}e^{3ik/2}.
\end{eqnarray}
Substituting these wave functions into Eqs. (\ref{eq:appendix1}) 
and (\ref{eq:appendix2}) with $k=\pi/2$ for $E=0$, 
one finds that the coefficient of the travelling wave in channel
$b$ is
\begin{equation} 
\tilde{T}=\frac{2t_2|w_{33}|}{||w_{33}w_{44}|+t_2^2|^2}. 
\end{equation}
Since the incoming flux and the outgoing flux are proportional to 
$|w_{33}|$ and $|w_{44}||\tilde{T}|^2$, respectively, the transmission factor from 
the left reservoir to the right is  
\begin{equation}
T_{ab}=\left|\frac{w_{44}}{w_{33}}\right|\,|\tilde{T}|^2
=\frac{4t_2^2|w_{33}w_{44}|}{||w_{33}w_{44}|+t_2^2|^2}. 
\end{equation}
Noticing $|w_{33}w_{44}|=-w_{33}w_{44}$, this coincides with the second 
fractional factor in Eq. (\ref{br_approx2}).

\end{document}